\documentclass[aps,twocolumn,prl,showpacs]{revtex4}
\usepackage[dvips]{graphicx}
\begin{document}

\title{Orbital Excitations in CMR Progenitor of LaMnO$_3$ studied by
  Resonant Inelastic X-ray Scattering} 
\author{T.~Inami$^{1,6}$,
  S.~Ishihara$^{2,6}$, H.~Kondo$^{3}$, J.~Mizuki$^{1,6}$,
  T.~Fukuda$^{1,6}$, S.~Maekawa$^{3,6}$, H.~Nakao$^{4,6}$,
  T.~Matsumura$^{5,6}$, K.~Hirota$^{5,6}$, Y.~Murakami$^{5,6}$, and
  Y.~Endoh$^{3,6}$ } 
\affiliation{$^{1}$ Synchrotron Radiation
  Research Center, Japan Atomic Energy Research Institute (SPring-8),
  Mikazuki, Hyogo 679-5148, Japan} 
\affiliation{$^{2}$Department of
  Applied Physics, University of Tokyo, Tokyo 113-8656 Japan}
\affiliation{$^{3}$Institute for Materials Research, Tohoku
  University, Sendai 980-8577, Japan} 
\affiliation{$^{4}$ Photon
  Factory, Institute of Materials Structure Science, High Energy
  Accelerator Research Organization, Tsukuba 305-0801, Japan}
\affiliation{$^{5}$Department of Physics, Tohoku University, Sendai
  980-8578, Japan} 
\affiliation{$^{6}$Core Research for Evolutional
  Science and Technology (CREST), Tsukuba 305-0047, Japan}

\date{received\hspace*{3cm}}

\begin{abstract}
  We report resonant inelastic x-ray scattering experiments of the
  orbital ordered manganite LaMnO$_3$.  When the incident photon
  energy is tuned near the Mn $K$ absorption edge, the spectra reveal
  three features at 2.5 eV, 8 eV and 11 eV.  The 8 eV and 11 eV peaks
  are considered charge-transfer type excitations.  On the other hand,
  theoretical calculations identify the 2.5 eV peak as an orbital
  excitation across the Mott gap, {\it i.e.} an electron excitation
  from the lower Hubbard band with the $d_{3x^2-r^2}$ and
  $d_{3y^2-r^2}$ orbital characters to the upper Hubbard band with the
  $d_{y^2-z^2}$ and $d_{z^2-x^2}$ ones.  The observed weak dispersion
  and characteristic azimuthal angle dependence of this new type of
  excitations are well reproduced by the theory which includes orbital
  degeneracy and strong electron correlation.
\end{abstract}


\maketitle

The interplay of three degrees of freedom of an electron, {\it i.e.},
spin, charge and orbital is now a central issue of the perovskite-type
manganites showing the colossal magnetoresistance (CMR) which should
be potential functional metallic materials in the next generation
\cite{tokura1}.  LaMnO$_3$ of the CMR progenitor is a Mott insulator
where the electronic configuration of the Mn$^{3+}$ ions is $t_{2g}^3
e_g^1$ with spin quantum number $S=2$. The band gap appears between
two $e_g$ bands of the Mn ions hybridized with the O $2p$ orbitals.
The occupied $e_g$ orbitals of $3d_{3x^2-r^2}$ and $3d_{3y^2-r^2}$ are
alternately ordered in the $ab$ plane accompanied by the lattice
distortion below 780K \cite{murakami2}.  With doping of holes in this
compound, a variety of electronic phases appear due to the mutual
interactions among the multi-degrees of freedom.  It is regarded, in
fact, that the CMR results from the collapse of a subtle balance under
the applied magnetic field destroying the orbital ordered state
\cite{tokura2,kaplan}.  Most of the previous explorations on mutual
interactions have been directed to spin and lattice dynamics by using
various microscopic probes such as neutrons, photons and electrons.

In contrast, little has been explored on dynamics of the orbital
degree of freedom.  In orbital ordered insulators, the highest
occupied and lowest unoccupied electronic states have different
orbital characters from each other.  Thus, an electron-hole excitation
across the Mott gap changes the symmetry of electronic cloud; this is
an orbital excitation.  It is well known that the virtual orbital
excitation between nearest-neighbor transition-metal ions brings about
the ferromagnetic superexchange interactions in insulators with
alternately ordered orbitals \cite{goodenough,kanamori}.  The observed
planar ferromagnetic spin alignment in LaMnO$_3$ signifies that this
excitation dominates the low-energy electronic excitation across the
Mott gap.  The knowledge of the orbital excitations is essential to
understand the physics of manganites.  Unfortunately, the orbital does
not couple directly to most of experimental probes.  Therefore, a
nature of the orbital dynamics in a wide range of energy and momentum
remains to be uncovered, unlike the spin and lattice dynamics.

Recent advances in synchrotron radiation source have changed this
situation. We report an exploration of orbital dynamics by
high-resolution resonant inelastic x-ray scattering (RIXS).  This is
an energy and momentum resolving probe
\cite{platzman,cu-abbamonte,cu-john,cu-hamalainen,cu-hasan}.
The resonant effects in x-ray diffraction (resonant elastic x-ray
scattering) availed recently to detect the orbital ordering in a
number of correlated electron systems \cite{murakami1,murakami2}.  By
tuning the incident x-ray energy to the absorption edge, scattering
becomes dramatically sensitive to the microscopic anisotropy of the
electronic clouds.  Here, RIXS is applied to the CMR progenitor
LaMnO$_3$.  We observed a new peak around the energy transfer 2.5 eV in
the orbital ordered state.  Its momentum and polarization dependence is
entirely distinct from the previous reports on cuprates, which have no
orbital degree of freedom \cite{cu-hasan,tsutsui}, but is in accord
with a theory including the orbital degeneracy.  This new probe would
tell us the entire picture of the spin-charge-orbital complex system,
i.e. CMR manganites.

The experiments were carried out at beam line BL11XU of SPring-8.
Incident x rays from a SPring-8 standard undulator were
monochromatized by a double-crystal diamond (111) monochromator, and
were focused onto a sample by a horizontally bent mirror.  The photon
flux and the typical spot size at the sample position were about
$2\times10^{12}$ photons/sec and 0.12 mm(H) $\times$ 1.4 mm(V),
respectively.  The horizontally scattered photons were collected by a
spherically bent ($R$ = 2 m) Ge (531) crystal of diameter 76 mm.  The
total energy resolution, determined from the quasi-elastic scattering
from the sample, was about 0.5 eV full width at half maximum (FWHM)
\cite{ixs-inami}. Two single crystals of LaMnO$_3$ cut from the same
boule were used.  One has a [110] direction parallel to the surface
normal ($Pbnm$ setting).  The other consists of two domains due to
twinning, and has $a$ and $b$ axes parallel to the surface normal.
All data were taken at room temperature.

\begin{figure}[b]
\resizebox{0.6\columnwidth}{!}{\includegraphics{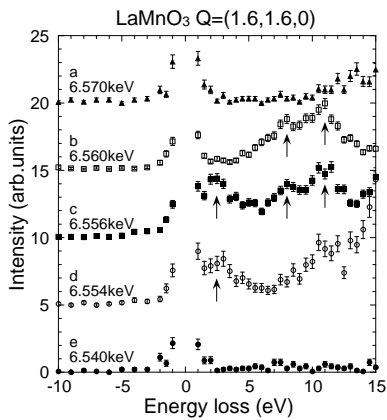}}
\caption{
  Inelastic x-ray scattering spectra of LaMnO$_3$ at $\vec
  q$=(1.6,1.6,0) for five incident energies near the Mn $K$ edge.
  Data are offset vertically for clarity. Three features at 2.5 eV, 8
  eV and 11 eV are shown by arrows.}
\label{fig:e-dep} 
\end{figure}

The inelastic scattering is plotted in Fig.\ref{fig:e-dep} as a
function of energy loss for several incident energies ($E_{\rm i}$)
across the Mn $K$ absorption edge (= 6.553 keV).  From these spectra,
resonant nature of the scattering is obvious.  Well below and above
the absorption edge (spectra a and e), no feature is observed except
the elastic scattering at zero energy transfer; the scattering is
observed only when the incident energy is close to the absorption
edge.  The salient spectral features are three peaks at 2.5 eV, 8 eV
and 11 eV, as seen in the spectrum c.  Note that the resonant energies
of the peaks are different from each other.  Only the excitation at
2.5 eV is observed in the spectrum d, in addition to the broad
scattering of the Mn $K\beta_5$ emission line ($3d\rightarrow1s$).  On
the other hand, the 2.5 eV peak vanishes in the spectrum b, while the
8 eV and 11 eV peaks remain.  This implies different origins of these
three peaks.

\begin{figure}[b]
\resizebox{1.0\columnwidth}{!}{\includegraphics{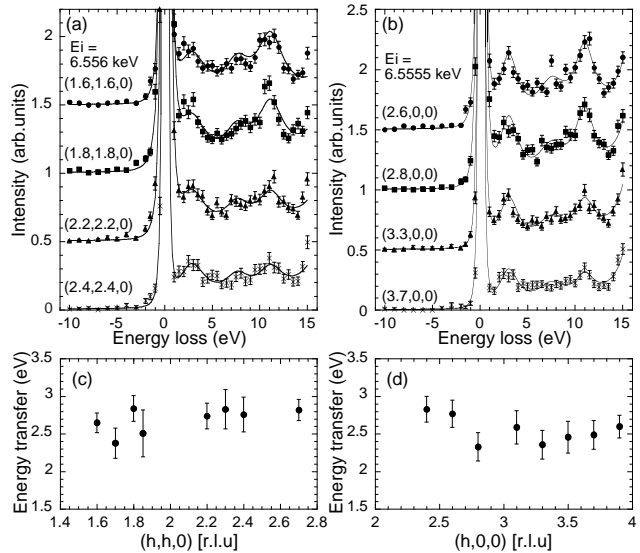}}
\caption{
  $q$-dependence of the resonant x-ray scattering spectra along ${\vec
    q}= (h,h,0)$ (a) and ${\vec q}= (h,0,0)$ (b) at incident photon
  energies 6.556 and 6.5555 keV, respectively.  Data are offset for
  clarity.  Solid lines are guides to the eye.  The peak positions of
  the 2.5 eV peak obtained from the fitting are also shown ((c) and
  (d)).}
\label{fig:q-dep} 
\end{figure}

The momentum ($q$) dependence of the inelastic scattering along ${\vec
  q}=(h,h,0)$ and $(h,0,0)$ is shown in Figs.\ref{fig:q-dep} (a) and
(b), respectively.  Three peaks centered at 2.5 eV, 8 eV and 11 eV
seem to show rather flat dispersion. The fact is shown more
convincingly by the following fitting procedure.  Utilizing the fact
that the resonant energy of the 2.5 eV peak differs from that of the 8
eV and 11 eV peaks, we extracted low-energy spectra by subtracting
data obtained at higher incident energies (6.560 keV for $(h,h,0)$
and 6.5585 keV for $(h,0,0)$) from those obtained at lower incident
energies (6.556 keV for $(h,h,0)$ and 6.5555 keV for $(h,0,0)$), and
fitted them by a Lorentzian curve.  The energy width was fixed in the
fitting. The peak position of the 2.5 eV peak obtained is plotted in
Figs.\ref{fig:q-dep} (c) and (d).  The energy dispersion of the 2.5 eV
peak is less than a few hundred meV, which should be reexamined by the
future measurement with higher energy resolution. We also fitted data
measured at $E_{\rm i}$ = 6.560 keV along ${\vec q}=(h,h,0)$ for the 8
eV and 11 eV peaks.  Data were well reproduced by two Lorentzian
curves with $q$ independent peak position and line width. The
intensity of the 8 eV peak shows weak $q$ dependence, while the
intensity of the 11 eV peak largely decreases as $q$ increases.

The 8 eV and 11 eV peaks are considered charge-transfer (CT) type
excitations.  The 8 eV peak is probably a transition from the O $2p$
orbitals to the unoccupied Mn $3d$ orbitals.  In previous RIXS studies
on cuprates, CT excitations are observed at about 6 eV.  In addition
to the similarity in the excitation energy, the weak $q$ dependence of
the 8 eV peak in peak position and in integrated intensity agrees well
with the reported results about the CT excitation of cuprates
\cite{cu-john,cu-abbamonte,cu-hamalainen}.  On the other hand, optical
conductivity measurements showed that a broad peak exists around
10~eV, which is attributed to transition from the O $2p$ orbitals to
the Mn 4$s$ or 4$p$ orbitals \cite{lmo-arima}.  The 11 eV peak can be
ascribed to this transition.

The lowest newly found peak around 2.5eV can be understood in terms of
the orbital excitation across the Mott gap, i.e., the electron-hole
excitation from the effective lower Hubbard band (LHB) to the upper
Hubbard band (UHB).  LaMnO$_3$ is identified as a CT insulator
\cite{saitoh2}; there exist strong on-site Coulomb interactions in the
Mn ion.  The insulating gap is formed between UHB consisting of the
unoccupied $3d$ orbital and the effective LHB consisting of the O $2p$
orbitals which are strongly mixed with the occupied Mn $3d$ ones.  In
the orbital-ordered phase, LHB has a symmetry of the $3d_{3x^2-r^2}$
and $3d_{3y^2-r^2}$ orbitals, and UHB has its counterpart, i.e. that
of the $3d_{y^2-z^2}$ and $3d_{z^2-x^2}$ orbitals.  The excitations
from LHB to UHB across the Mott gap (or the CT gap) change the
symmetry of the electron wave function.

\begin{figure}[t]
\resizebox{0.47\columnwidth}{!}
{\includegraphics*[50mm,110mm][160mm,260mm]{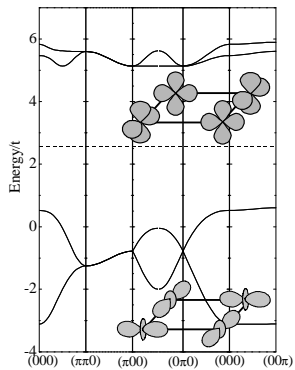}}
\caption{
  The electronic band structure for the orbital ordered LaMnO$_3$.
  The Hartree-Fock approximation is applied to the generalized Hubbard
  model where two $e_g$ orbitals and intra-site Coulomb interactions
  are taken into account.  The alternate ordering of the
  $3d_{3x^2-r^2}$ and $3d_{3y^2-r^2}$ orbitals in the $ab$ plane is
  assumed.  Broken line indicates the chemical potential located at
  the center of the highest occupied and lowest unoccupied bands.  $t$
  denotes the hopping integral of electrons between
  nearest-neighboring $3d_{3z^2-r^2}$ orbitals in the $z$ direction
  and is about $0.5-0.7$~eV.  The inset shows schematic illustration
  of the orbital ordered state; the $3d_{y^2-z^2}$ and $3d_{z^2-x^2}$
  orbitals (top) constituting the upper Hubbard band and the
  $3d_{3x^2-r^2}$ and $3d_{3y^2-r^2}$ orbitals (bottom) constituting
  the lower Hubbard band.  }
  \label{fig:theory} 
\end{figure}

A theory of RIXS from the orbital excitations is developed on the
generalized Hubbard model with orbital degeneracy.  Consider the
scattering of the incident x rays with momentum $\vec k_i$, energy
$\omega_i$ and polarization $\lambda_i $ to $\vec k_f$, $\omega_f$ and
$\lambda_f$.  The x-ray scattering cross section \cite{bulme,ma} is
formulated by the Liouville-operator method as \cite{kondo}:
\begin{eqnarray}
{d^2 \sigma \over d \Omega d \omega_f}&=& \sigma_e
{\omega_f \over \omega_i}
\sum_{\alpha, \alpha'=x,y,z}
(\vec e_{ k_f \lambda_f})_{\alpha}    (\vec e_{k_i \lambda_i})_{\alpha}
\nonumber \\
&\times&
(\vec e_{ k_f \lambda_f})_{\alpha'}   (\vec e_{k_i \lambda_i})_{\alpha'}
\Pi_{\alpha' \alpha}(\omega_i-\omega_f, \vec k_i-\vec k_f) ,
\label{eq:sigma2}
\end{eqnarray}
where $\sigma_e=( e^2 / mc^2)^2$ and $\vec e_{k \lambda}$ is the polarization of the x rays. 
The Fourier transform of
$\Pi_{\alpha' \alpha}(\omega_i-\omega_f, \vec k_i-\vec k_f)$
is given by
\begin{equation}
\Pi_{\alpha' \alpha}(t,\vec r_{l'}-\vec r_{l})=
{|B|^4 \over m^2} \sum_{\gamma \gamma'} 
D_{\gamma'  \alpha'}^\ast D_{\gamma \alpha}
\langle T_{l' x}(t) T_{l x} (0) \rangle , 
\label{eq:corrr}
\end{equation}
with a constant $B$ 
and a factor $D_{\gamma \alpha}$ which gives the
amplitude of the orbital excitations from the $3d_{\gamma}$ orbital by x ray with polarization $\alpha$. 
The orbital degree of freedom is represented by the pseudo-spin operator 
$\vec T_l$ with quantum number $T=1/2$. 
$T_z=\pm 1/2$ corresponds to the state where one of the two orbitals 
is occupied by an electron. 
Thus, $T_x( \equiv {1 \over 2}(T_++T_-))$ in Eq.~(\ref{eq:corrr}) 
implies the orbital excitation. 
The Jahn-Teller type lattice distortion is introduced in the model  
and the adiabatic approximation is adopted. 
This is because the RIXS process is supposed to be faster than the lattice relaxation time.  
The detailed formulation was presented in Ref.~\onlinecite{kondo}.
Note that the cross section is represented by the dynamical
correlation function of the pseudo-spin operators of orbital.  
The obtained spectra show a gap
about $4t$ and a broad peak centered around $6t$, where $t$ is the
hopping integral between the nearest-neighboring $3d_{3z^2-r^2}$
orbitals in the $z$ direction and is about $0.5-0.7$~eV.  The
calculated spectra are fitted by a Lorentzian curve.
The obtained
dispersion of the center of the curve 
exhibits a weak momentum
dependence within 0.1$t$.  This almost flat dispersion
is attributed to the effects of orbital order.  The RIXS spectra are
approximately given by the convolution of LHB with the $3d_{3x^2-r^2}$
and $3d_{3y^2-r^2}$ orbital characters and UHB with the $3d_{y^2-z^2}$
and $3d_{z^2-x^2}$ ones (see Fig.\ref{fig:theory}).  Since the
electron hopping between the unoccupied orbitals is forbidden in the
$ab$ plane, the dispersion of UHB is almost flat.  As a result, the
density of state of the LHB dominates the RIXS spectra.  This
characteristic of the theory based on the orbital excitations is
consistent with the experimental results shown in Fig.\ref{fig:q-dep}.

\begin{figure}[t]
\resizebox{1.0\columnwidth}{!}{\includegraphics{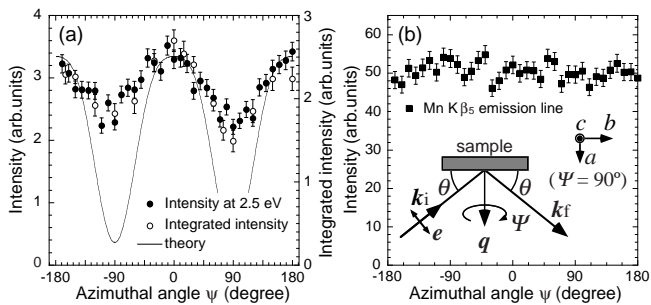}}
\caption{
  (a) Azimuthal angle ($\psi$) dependence of the intensity of the 2.5
  eV peak at $\vec q$=(3.4,0,0).  Open and solid circles show the
  integrated intensity of the 2.5 eV peak and the intensity at the
  energy transfer 2.5 eV, respectively.  A solid line shows the
  integrated intensity of the theoretically calculated RIXS spectra
  for the orbital ordered LaMnO$_3$ as a function of $\psi$.  The
  intensity is scaled so that the maxima approximately agree with the
  experimental data.
  (b) Azimuthal angle dependence of the intensity of the Mn $K\beta_5$
  emission line at $\vec q$=(3.4,0,0).  $E_{\rm i}=6.555$ keV and 
  $E_{\rm f}=6.5355$ keV.  The intensity is independent of $\psi$.
  Experimental setup for the azimuthal-angle dependence measurement is
  shown in inset.  The crystal is rotated about the scattering vector
  $\vec q$=(3.4,0,0).  $\theta$ is half the scattering angle and was
  about 34$^\circ$.  At $\psi=90$ and 0$^\circ$, the $c$ axis is
  perpendicular to and in the scattering plane, respectively. }
\label{fig:azi-dep} 
\end{figure}

The significance of the RIXS spectra in the orbital ordered state is
that the scattering intensity depends on the azimuthal angle, which is
the rotation of the sample about the scattering vector $\vec q$.  In
the resonant elastic x-ray scattering, this is a direct evidence of
the anisotropic scattering factor and is crucial for identifying the
scattering from ordered orbitals \cite{murakami1}.  The azimuthal
angle dependence was measured using the experimental setup shown in
Fig.\ref{fig:azi-dep}.  The polarization of the incident photon is in
the scattering plane ($\pi$ polarization).  When the $c$ axis is
perpendicular to the scattering plane, the azimuthal angle $\psi$ is
defined as 90$^\circ$.  The integrated intensity of the the 2.5 eV
peak at ${\vec q}=(3.4,0,0)$ is shown in Fig.\ref{fig:azi-dep}(a)
(open circles).  The 2.5 eV peak was extracted by subtracting an
inelastic spectrum measured at $E_{\rm i}$=6.560 keV from that
measured at $E_{\rm i}$=6.5555 keV, and was fitted by a Lorentzian
curve.  The peak position and the width were fixed at 2.4 eV and 3.0
eV (FWHM), respectively, in the fitting.  The intensity at the energy
transfer 2.5 eV at ${\vec q}=(3.4,0,0)$ is also shown by solid
circles.  Both intensities exhibit a characteristic oscillation with a
twofold symmetry and take its maxima at $\psi=0^\circ$ and
$180^\circ$.  This behavior is in qualitative agreement with the
theoretical calculation for the orbital excitation (solid line).  The
type of the orbital ordered state is the same with those adopted in
Fig.~\ref{fig:theory}.  A quantitative discrepancy in the oscillation
amplitude between the theory and data may be attributed to the
multi-domain nature of the sample and the mean-field approximation
utilized in the calculation.  The similar azimuthal angle dependence
was observed for the 8 eV and 11 eV peaks as well.  We also measured
the Mn $K \beta_5$ fluorescence line, of which intensity is
independent of the azimuthal angle (Fig.\ref{fig:azi-dep}(b)).  This
result illustrates that the observed azimuthal angle dependence does
not arise from extrinsic origins, such as anisotropic absorption.
Moreover, we would like to emphasize that the azimuthal angle
dependence of the 2.5 eV peak is not caused by the shift of the
resonant energy.  No significant change was observed in the resonant
energy of the 2.5 eV peak at $\psi=0$ and $90^\circ$, in sharp
contrast to the case of Nd$_2$CuO$_4$ \cite{cu-hamalainen}.
Theoretically, the azimuthal angle dependence originates from the fact
that an $x$-polarized photon does not induce the orbital excitations
in a site where the $3d_{3x^2-r^2}$ orbital is occupied, because the
local electronic symmetry at this site is not broken by the
$x$-polarized photon.  The observed azimuthal angle dependence
strongly supports the present interpretation that the 2.5 eV peak
originates from the orbital excitations.

Through the present experimental study of RIXS combined with the
theoretical calculation, we were able to identify the inelastic peak
at 2.5 eV in LaMnO$_3$ as the orbital excitations by its momentum and
polarization dependence.  As a good comparison with the present study,
the application to hole doped manganites is promising.  Simple
metallic behavior without orbital order will result in distinct
dispersion of low-energy features.  Another possible application of
RIXS is to detect the dispersion of collective orbital excitations
(orbital waves), which were glimpsed in the recent Raman scattering
experiments at zero momentum \cite{saitoh}.  As resonant x-ray
diffraction study has elucidated static order of orbital in transition
metal oxides, RIXS becomes a powerful tool for investigating dynamics
of orbital and related phenomena.

We thank T. Iwazumi and T. Arima for helpful discussions, and are
grateful to P. Abbamonte for making the diced Ge analyzer.  This work
was supported in part by Grant-in-Aid for Scientific Research Priority
Area from the Ministry of Education, Science, Sport, Culture and
Technology of Japan, and Science and Technology Special Coordination
Fund for Promoting Science and Technology of Japan.

\end{document}